\documentclass[preprint]{aastex} 

\newcommand{\be}{\begin{equation}}
\newcommand{\ee}{\end{equation}}
\newcommand{\bwig}{ {\widetilde b}} 
\newcommand{\tauwig}{ {\widetilde \tau}} 
\newcommand{\sigwig}{ {\widetilde \sigma}} 
\newcommand{\dwig}{ {\widetilde D}} 

\begin{document}

\title{Enhancement of Ambipolar Diffusion Rates \\
through Field Fluctuations}
\medskip
\author{Marco Fatuzzo$^1$ and Fred C. Adams$^{2,3}$}
\bigskip 
\affil{$^1$Physics Department, Xavier University, Cincinnati, OH 45207} 

\affil{$^2$Michigan Center for Theoretical Physics \\ 
Physics Department, University of Michigan, Ann Arbor, MI 48109}

\affil{$^3$Astronomy Department, University of Michigan, Ann Arbor, MI 48109}

\begin{abstract} 

Previous treatments of ambipolar diffusion in star-forming molecular
clouds do not consider the effects of fluctuations in the fluid fields
about their mean values. This paper generalizes the ambipolar 
diffusion problem in molecular cloud layers to include such
fluctuations. Because magnetic diffusion is a nonlinear process,
fluctuations can lead to an enhancement of the ambipolar diffusion
rate. In addition, the stochastic nature of the process makes the
ambipolar diffusion time take on a distribution of different values.
In this paper, we focus on the case of long wavelength fluctuations
and find that the rate of ambipolar diffusion increases by a
significant factor $\Lambda \sim 1 - 10$. The corresponding decrease
in the magnetic diffusion time helps make ambipolar diffusion more
consistent with observations.

\end{abstract}


\section{Introduction}          \label{sec:intro}

In the usual paradigm of low mass star formation, molecular cloud
cores are supported by magnetic fields. In order for star formation to
take place, the cores must lose magnetic support, and this loss of
support is generally thought to take place through the action of
ambipolar diffusion (Mouschovias 1976; Shu 1983; Nakano 1984; Shu,
Adams, \& Lizano 1987; Lizano \& Shu 1989; Ciolek \& Basu 2000,
2001). This general picture has support from observations, which
suggest that ion-neutral drift does indeed occur in magnetized
star-forming cores (e.g., Greaves \& Holland 1999).

An important issue facing this standard scenario is the time scale
required for magnetic support to be removed from the cloud cores. As
the observational picture comes into sharper focus, the number of
observed cores without stars (e.g., Jijina, Myers, \& Adams 1999)
seems to be smaller than that predicted by most previous estimates
from ambipolar diffusion (e.g., Ciolek \& Mouschovias 1994, 1995;
Lizano \& Shu 1989) by a factor of 3 -- 10. In other words, loss of
magnetic support by diffusion appears to be too slow, with a time
scale a factor of 3 -- 10 times longer than suggested by the observed
statistics of cloud cores. However, these previous calculations
neglected a dimensionless factor that depends on the mass to flux
ratio of the cores (Ciolek \& Basu 2001). If the cloud cores have mass
to flux ratios that approach the critical value, then the ambipolar
diffusion time scale is significantly shorter than previous estimates. 
In particular, if the mass to flux ratio becomes supercritical, then
the ambipolar diffusion time scale approaches zero. The need for this
correction is bolstered by a recent compilation of Zeeman measurements
of magnetic field strengths (Crutcher 1999), which suggests that many
cores may have mass to flux ratios near the supercritical value. This
observed sample includes only 27 cores with relatively large masses;
additional measurements are necessary to clarify the observational
picture.

In this study, we consider the effects of fluctuations on the
mechanism of ambipolar diffusion described above. The time scale issue
remains important and this work shows that ambipolar diffusion can
operate more quickly in the presence of such fluctuations. In
addition, because of the chaotic nature of the fluctuations, the
ambipolar diffusion time scale will take on a full distribution of
values for effectively ``the same'' initial states.

Fluctuations are expected to be present in essentially all star
forming regions. Molecular clouds are observed to have substantial
non-thermal contributions to the observed molecular line-widths (e.g.,
Larson 1981; Myers, Ladd, \& Fuller 1991; Myers \& Gammie 1999).
These non-thermal motions are generally interpreted as arising from
MHD turbulence (e.g., Arons \& Max 1975; Gammie \& Ostriker 1996; for
further evidence that the observed linewidths are magnetic in origin,
see Mouschovias \& Psaltis 1995).  Indeed, the size of these
non-thermal motions, as indicated by the observed line-widths, are
consistent with the magnitude of the Alfv{\'e}n speed (e.g., Myers \&
Goodman 1988; Crutcher 1998, 1999; McKee \& Zweibel 1995; Fatuzzo \&
Adams 1993). As a result, the fluctuations are often comparable in
magnitude to the mean values of the fields (T. Troland, private
communication).

Background fluctuations can lead to a net change in the diffusion rate
because magnetic diffusion is a nonlinear process. As many authors
have derived previously (e.g., see the textbook treatment of Shu
1992), and as we present below, the (dimensionless) diffusion 
equation takes the schematic form 
\be 
{\partial b \over \partial \tau} = {\partial \over \partial \mu} 
\Bigl( b^2 {\partial b \over \partial \mu} \Bigr) \, ,
\label{eq:magone} 
\ee 
where $b$ is the magnetic field strength, $\mu$ is the Lagrangian mass
coordinate, and we have ignored density variations.  Now suppose that
the magnetic field fluctuates about its mean value on a time scale
that is short compared to the diffusion time (the time required for
the mean value to change). We thus let $b \to b(1+\xi)$, where $\xi$
is the relative fluctuation amplitude. In the simplest case in which
the fluctuations are spatially independent, the right hand side of
equation [\ref{eq:magone}] is thus multiplied by a cubic factor $(1 +
\xi)^3$. Although a linear correction would average out over time,
this nonlinear term must always average to a value greater than unity
and the corresponding diffusion time scale grows shorter by the same
factor. As an example, suppose the field spends half of its time at a
value of twice its mean strength and the other half of its time near
zero strength. Half of the time, the effective diffusion constant is
thus larger by a factor 8, whereas the other half of the time, the
diffusion constant is effectively zero. In this naive example, the
mean diffusion constant is thus 4 times larger due to the
fluctuations. The goal of this paper is to derive a more rigorous
argument for this time scale enhancement.

This effect -- changing diffusion time scales because of fluctuations
-- is well known in mathematical subfields. Simpler problems in which
random noise fields drive physical systems at different rates appear
in a host of textbooks (e.g., Srinivasan \& Vasudevan 1971; Soong
1973).  More recently, in the context of ``stochastic ratchets'', it
has been shown that random fluctuations can drive a physical system to
propagate ``uphill'', i.e., the opposite direction of its natural
propagation in the absence of fluctuations (Doering, Horsthemke, \&
Riordan 1994). In astrophysics, stochastic aspects of magnetic field
fluctuations have been considered in the context of cosmic ray
propagation (e.g., Jokipii \& Parker 1969) and also in stellar
atmospheres (e.g., Shore \& Adelman 1976). Several previous papers
have studied turbulent fluctuations in magnetically supported clouds,
often by considering how the turbulence itself leads to field
evolution in the absence of ambipolar diffusion (e.g., Kim 1997). The
formation of cores through the dissipation of turbulence has been
suggested (Myers \& Lazarian 1998). Turbulence can also enhance the
rate of ambipolar drift and may help explain the observed relationship
between density and magnetic field strength, $B \propto \rho^\kappa$
(see Zweibel 2001). In this current work, we consider ambipolar
diffusion to be the main process that forms molecular cloud cores and
study how fluctuations alter its rate.

This paper is organized as follows. In \S 2, we reformulate the
ambipolar diffusion calculation in a plane geometry, where we
explicitly include fluctuations in both the magnetic field and the
density field. We perform an analysis of the resulting set of
equations in \S 3. We specialize to the limit of long wavelength
fluctuations and apply the resulting formalism to astrophysical
systems. We conclude, in \S 4, with a summary and discussion of our
results. The case of short wavelength fluctuations is presented
briefly in an Appendix. The most important outcome of this study is 
to demonstrate that fluctuations can lead to more rapid diffusion of
magnetic fields in star forming regions and that the diffusion time
scale takes on a distribution of values (rather than a single time
scale).

\section{Ambipolar Diffusion in the Presence of Fluctuations} 

In this section, we modify the standard ambipolar diffusion derivation
to include the effects of fluctuations. We consider the simplest case
of a planar layer of molecular cloud material. The magnetic field
lines are parallel to the plane and all quantities depend only on the
height $z$ above the midplane, 
\be
{\bf B} = B(z) \, {\hat x} \, . 
\ee
We note that a host of two-dimensional axisymmetric calculations have
already been done (Lizano \& Shu 1989; Mouschovias \& Morton 1991;
Ciolek \& Mouschovias 1994; Basu \& Mouschovias 1994; Basu 1997, 1998;
Ciolek \& Basu 2000).  Although more idealized, this one-dimensional
calculation is useful because it allows for analytical results and
isolates the fluctuation effects that we are trying to elucidate 
(see also the discussion at the end of this section). 

As derived previously (e.g., Shu 1992), with this basic configuration
the equations of motion take the form 
\be
{\partial \rho \over \partial t} + 
{\partial \over \partial z} (\rho u) = 0 \, , 
\ee
\be
{\partial u \over \partial t} + u {\partial u \over \partial z} = 
g - {1 \over \rho} 
{\partial \over \partial z} \Bigl( P + 
{B^2 \over 8 \pi} \Bigr) \, , 
\ee
\be
{\partial g \over \partial z} = - 4 \pi G \rho \, , 
\ee
and finally 
\be
{\partial B \over \partial t} + 
{\partial \over \partial z} (B u) = 
{\partial \over \partial z} \Bigl( 
{B^2 \over 4 \pi \gamma \rho \rho_i} 
{\partial B \over \partial z} \Bigr) 
\, . 
\ee
To close the system of equations, the ion mass density $\rho_i$ must be
specified. The ion population depends on the complex balance between
the ionization rate of the neutrals (primarily through cosmic rays), the
subsequent production of molecular and metal ions, and the ion-electron 
recombination rate in the presence of grains. A reasonable (standard) 
approximation for the ionic mass density in molecular cloud environments 
is given by
\be
\rho_i  = C \rho^{1/2}\;,
\label{eq:rhoion} 
\ee
where $C = 3\times 10^{-16}$ cm$^{-3/2}$ g$^{1/2}$. However, a new
observational study (Caselli et al. 2001) indicates that the
ionization fraction can be significantly smaller than suggested by
equation [\ref{eq:rhoion}] with this value of the constant $C$ (see
also Ciolek \& Mouschovias 1998).  This reduction in ionization also
acts to speed up the ambipolar diffusion rate and helps alleviate the
time scale problem of interest.  Notice, however, that this effect is
independent of our present work and can be easily incorporated by 
using a different value of the constant $C$. 

Next, we introduce the fluctuations through the ansatz 
\be
B \to B (1 + \xi) \qquad {\rm and} \qquad 
\rho \to \rho (1 + \eta) \, . 
\label{eq:ans} 
\ee
The relative fluctuations $\xi$ and $\eta$ obey distributions, which
ultimately determine the effects of these variations. Although we will
specialize to particular distributions later on, we now keep the
analysis as general as possible. In any case, the fluctuations are
assumed to have zero mean so that the quantities $B$ and $\rho$
appearing in the equations of motion can be thought of as
time-averaged quantities and play the same role in this generalized
calculation as they do in previous treatments.

Following many previous treatments, we rewrite the problem in terms 
of Lagrangian coordinates so that the basic variable is the surface  
density of neutrals between the midplane ($z=0$) and a height
$z$. This change of variables takes the form 
\be
\sigma \equiv \int_0^z \rho (z', t) dz' \, .
\ee
Notice that this Lagrangian coordinate measures the distance in 
terms of the mean density $\rho$ rather than the full fluctuating 
density field $\rho (1+\eta)$. In terms of this new coordinate, 
the equation of continuity becomes 
\be
{\partial z \over \partial \sigma} = {1 \over \rho} \, . 
\label{eq:continuity} 
\ee
For an isothermal equation of state ($P = a^2 \rho$), 
the force equation can be written in the form 
\be
- {\partial^2 z \over \partial t^2} = 4 \pi G \int_0^z
\rho (1 + \eta) dz' + {a^2\over 1 + \eta} 
{\partial \over \partial \sigma} \bigl[ \rho (1 + \eta) \bigr] 
+ {1\over 1+\eta} {\partial \over \partial \sigma} 
\Bigl[ {B^2 (1 + \xi)^2 \over 8 \pi} \Bigr] \, . 
\ee
Since ambipolar diffusion is generally much slower than gravitational
collapse, the cloud is expected to evolve near a magnetohydrodynamic
equilibrium state.  The inertial term in the force equation can
therefore be justifiably ignored, leaving an expression which can be
integrated to yield the quasi-magnetohydrodynamic equilibrium
condition. In this case, however, the integration can only be carried
out after a suitable averaging of the fluctuations has been done
(see below). Finally, the magnetic field evolves according to a 
nonlinear diffusion equation of the form 
\be
{\partial \over \partial t} \Bigl[
{B (1 + \xi) \over \rho (1 + \eta)} \Bigr] 
= {1 \over 1 + \eta} {\partial \over \partial \sigma} 
\Biggl\{ {B^2 (1 + \xi)^2 \over 4 \pi \gamma C \rho^{1/2} 
(1 + \eta)^{3/2}} {\partial \over \partial \sigma} \bigl[ 
B (1 + \xi) \bigr] \Biggr\} \, ,
\label{eq:diffuse0} 
\ee 
where the partial derivatives are taken with respect to 
constant $t$ or constant $\sigma$. 

Following the notation of Shu (1983, 1992), we introduce a
dimensionless surface density $\mu$, volume density $p$, magnetic
field $b$, vertical coordinate $y$, and time $\tau$ as follows: 
\be
\sigma \equiv \sigma_\infty \mu\, ,
\label{eq:ndsigma} 
\ee
\be
\rho \equiv {2\pi G \sigma_\infty^2\over a^2} p \, ,
\label{eq:ndrho} 
\ee
\be
B \equiv 4\pi G^{1/2} \sigma_\infty b\, ,
\label{eq:ndmag} 
\ee
\be
z \equiv {a^2\over 2\pi G \sigma_\infty} y\, ,
\label{ndspace} 
\ee
and 
\be
t \equiv \left[{\gamma C\over 2 (2\pi G)^{1/2}}\right]
\left[{a\over 2\pi G \sigma_\infty}\right] \tau\, . 
\label{eq:ndtime} 
\ee
The closed set of equations which describe ambipolar diffusion can 
now be written as a continuity equation 
\be
{\partial y \over\partial \mu} = {1\over p}\, , 
\label{eq:contin}
\ee
a force balance equation 
\be
\int_0^y p (1 + \eta) dy' + {1\over 1+\eta}
{\partial\over\partial\mu} \bigl[ p (1+\eta) \bigr] 
+ {1\over 1+\eta} {\partial\over\partial\mu}
\Bigl[ b^2 (1+\xi)^2 \Bigr] \, = 0 \, , 
\label{eq:force}
\ee
and a magnetic diffusion equation 
\be
{\partial\over\partial\tau} 
\left[ {b (1+\xi) \over p (1+\eta)} \right] = 
{1 \over 1+\eta} {\partial \over \partial\mu} 
\left\{ {b^2(1+\xi)^2\over p^{1/2} (1+\eta)^{3/2}} 
{\partial \over \partial\mu} [b(1+\xi)] \right\} \, . 
\label{eq:diffuse} 
\ee
These three equations, in conjuntion with the specification 
of the fluctuations, constitute the problem to be solved. 

As is well known, this one-dimensional slab model is unrealistic in
the sense that gravity ``saturates''. In other words, in Lagrangian
coordinates, our model has a constant gravitational field strength for
a given column density.  The gravitational field depends on the mass
shell but not on the distance from the midplane. Hence, a Lagrangian
observer would find no increase in the gravitational field strength as
the slab is compressed, unlike the more realistic case of higher
dimensions. This point is important because ambipolar diffusion is
driven by self-gravity and the slab model lies in the regime of ``weak
gravity'' (Mouschovias 1982). However, the enhancement of the time
scale that we study here arises from the nonlinearity of the equations
and will be present in any geometry. To illustrate this claim, we
derive an analogous formulation for a cylindrical geometry in Appendix
A. In particular, we find that the enhancement factor (due to
fluctuations) takes the same form as that derived above. Although the
absolute value of the ambipolar diffusion time scale depends on the
geometry, the enhancement factor calculated here does not.

\section{Analysis} 

In this section, we find solutions to a limited version of the problem
derived in the previous section.  Unfortunately, a full solution to
equations [\ref{eq:contin} -- \ref{eq:diffuse}] is beyond the scope of
this paper and several important simplifications are necessary to make
further progress. 

In a complete theory, one would also derive equations of motion for
the fluctuations, represented here by $\xi$ and $\eta$, and one would
solve for their temporal and spatial dependences in a self-consistent
manner. Such a calculation would require a theory of how MHD waves are
produced, and how they cascade into turbulence.  Because we have no
working {\it a priori} theory of turbulence, however, our first
simplification is to assume viable fluctuation distributions rather
than calculate them. In other words, we adopt a semi-empirical
approach. These fluctuations are observed, and we can use observations
to constrain their form, but we do not explicitly calculate their
behavior.

Regardless of what distributions are chosen for the fluctuations,
another complication arises. The equations of motion derived above
describe the evolution of one particular realization of the
problem. The usual way to solve such a problem is to set up the
initial state and step forward in time (either analytically or
numerically) by sampling the values of the fluctuations $\xi$ and
$\eta$ from their (presumably known) distributions.  As time
progresses, the field evolves and one obtains a solution. However,
this approach -- which we will adopt -- does not provide a full
solution to the problem. Because the fluctuations obey a distribution,
a probabilistic description of the overall evolution problem is
necessary. In particular, the time evolution of the magnetic field
will have a {\it distribution} of possible solutions.  A full solution
to the entire problem (see Doering 1990) would calculate the full
distribution of possible time behaviors for the magnetic field (and
other fluid variables). Because this issue is compounded for the case
of short wavelength fluctuations, we focus our analysis on long
wavelength fluctuations. We briefly discuss the case of short
wavelength fluctuations in Appendix B, which shows that additional
simplifying assumptions are necessary to make progress.

\subsection{Time Averaging} 

In order to describe the average behavior of the fluid fields, we
introduce a method of intermediate time averaging.  Previous
calculations for slab models suggest that ambipolar diffusion occurs
on a time scale $\tau_{AD} \sim 10 - 15$ (Nakano 1984, Shu 1983,
Mouschovias 1983), although this time scale is significantly shorter
when the mass to flux ratio of the cores approach supercritical values
(Ciolek \& Basu 2001). For comparison, observations suggest that the
real value is closer to $\tau_{AD} \sim 1 - 3$. The fluctuations
themselves presumably occur within the MHD regime and thus occur on
time scales $\tau_{\rm mhd} \sim \lambda / v_A$, where $\lambda$ is
the length scale of the fluctuation and $v_A$ is the Alfv\'en speed.
We can write the length scale in dimensionless form through the
relation 
\be
\lambda = {a^2\over 2\pi G\sigma_\infty} \, \chi \, , 
\label{eq:chidef} 
\ee
where we expect $\chi \approx 1 - 10$ for the long wavelength limit 
considered below (see \S 3.3). The dimensionless time scale for which 
the fluctuations vary is thus given by 
\be
\tau_{\rm mhd} \sim 
\left[{\gamma C\over 2(2\pi G)^{1/2}}\right]^{-1}
\,{a\over v_A}\,\chi\, , 
\label{eq:taumhd} 
\ee 
where the dimensionless parameter $[\gamma C / 2(2\pi G)^{1/2}]$
$\sim$ 8 for our adopted values of $C$ and $\gamma$. For this approach
to be consistent, the fluctuation time scale must be much shorter than
the ambipolar diffusion time.  This requirement implies a constraint
of the form $\chi \ll 8 (v_A / a) \tau_{AD}$, where we have used
standard values for $\gamma$ and $C$ (see also Zweibel 1988). Notice
that the ratio $v_A / a$ is typically 5 -- 10, so this constraint is
usually easy to satisfy (see, however, the discussion below).

To find time-averaged quantities, we must average over 
an intermediate time scale $\tau_0$ that obeys the ordering 
\be
\tau_{\rm mhd} \ll \tau_0 \ll \tau_{AD} \, . 
\ee
Again, we require $\tau_{\rm mhd} \ll \tau_{AD}$ so that an
intermediate range of time scales exists.  Further, we let brackets
$\langle \dots \rangle$ denote time averaged quantities so that 
\be
\langle p (1+\eta) \rangle \equiv {1 \over \tau_0} 
\int_0^{\tau_0} p (1+\eta) d\tau' \approx p \, , 
\ee
and
\be
\langle b (1+\xi) \rangle \equiv {1 \over \tau_0} 
\int_0^{\tau_0} b (1+\xi) d\tau' \approx b \, .
\ee
In equating the time averaged quantities with $p$ and $b$, we are
ignoring errors of ${\cal O}$ $[\tau_0 / \tau_{AD}]$.  For the
reasonable choice of taking the averaging time scale to be the
geometric mean $\tau_0 = \sqrt{\tau_{\rm mhd} \tau_{AD}}$, the
relative error becomes ${\cal O}$ [$(\tau_{\rm mhd} /
\tau_{AD})^{1/2}$].

\subsection{The Quasi-Equilibrium State} 

Next, we need to consider the equilibrium states. The time scale over
which fluctuations change (roughly given by the MHD crossing time) is
much shorter than the collapse time scale (roughly given by the sound
crossing time) for cloud cores undergoing ambipolar diffusion. Because
the fluctuation time scale is also much shorter than the ambipolar
diffusion time scale, the slab is expected to evolve in a quasi-static
equilibrium state supported primarily by magnetic pressure. Since the
force resulting from magnetic pressure is nonlinear, however, the 
fluctuations can play an important role in supporting this state. The
quasi-equilibrium condition is found by integrating the time-averaged
force equation [\ref{eq:force}] to obtain 
\be
K b^2 + p = 1 - \mu^2 \qquad {\rm where} \qquad 
K \equiv \left<{(1+\xi)^2\over (1+\eta)}\right> \, .
\label{eq:kdef} 
\ee
In this geometry (used here to make the problem more tractable), an
equilibrium state exists even in the absence of magnetic fields and
full gravitational collapse is not possible.  While this geometry does
not reflect realistic conditions in molecular cloud cores, it nicely 
illustrates the effects of fluctuations on the ambipolar diffusion
problem.

In obtaining the above form, we have assumed that the fluctuations are 
relatively well-behaved so that they obey constraints of the form 
\be
\left< {1 \over 1 + \eta} {\partial\eta\over\partial\mu} \right> 
\approx 0 \qquad {\rm and} \qquad \left< (1 + \xi)  
{\partial \xi \over\partial\mu} \right> \approx 0 \, . 
\ee
In order for these constraints to hold, the derivatives of the
fluctuations must average to zero (as expected) {\it and} the
derivatives must not be correlated with the fluctuations themselves.
In other words, the fluctuations must be both spatially and temporally
symmetric.  Ambipolar diffusion, as expressed by equation
[\ref{eq:diffuse}], then evolves the quasi-equilibrium state from a
magnetically supported configuration to a thermally supported one in
the slab model.

%This set of equations can be solved to yield the time scale
%under which ambipolar diffusion operates.  

\subsection{Long Wavelength Fluctuations} 

We now consider the effects of fluctuations on the ambipolar diffusion
process. As shown below, these effects vary considerably with the
spectrum of the fluctuations. To start, we assume that the fluctuations 
in density and the magnetic field are independent of each other. We
also consider the simplest case in which the fluctuations have long
wavelengths. In the limiting case where the fluctuation scale
$\lambda$ is much larger than the typical length scale $R$ in the
cloud, the fluctuations can be considered to be spatially independent
in the diffusion equation [\ref{eq:diffuse}]. This regime of parameter
space is defined by the constraint 
\be
\lambda \gg R \sim {a^2 \over 2 \pi G \sigma_\infty} \, . 
\label{eq:spacecon} 
\ee
For the rest of this paper, we will specialize our analysis to long
wavelength fluctuations that obey this constraint. 

Although nature can also support short wavelength fluctuations, a
rigorous treatment of their effects is made difficult by several
issues: (1) In stochastic differential equations, the solutions depend
on the manner in which various limits are taken (see Doering 1990).
For short wavelength fluctuations, many different equivalent ways of
taking the appropriate limits are possible and no unique solution
exists without further specification. In this context, we would need
to understand the origin of turbulent fluctuations to provide further
specification.  (2) In numerical treatments of the diffusion problem,
the diffusion equation can be unstable on short length scales. Many
numerical treatments allow for (small) errors on short size scales,
but provide the correct global behavior (see Press et al. 1986). To
adequately follow short wavelength fluctuations, one needs a numerical
method that adequately resolves all spatial scales. In light of these
difficulties, we focus on long wavelength perturbations in this
paper. For purposes of illustration, however, we consider a
representative approach for short wavelength fluctuations in 
Appendix B.

In order for the fluctuations to affect the diffusion process, the
effective diffusion coefficient must sample a range of possible
fluctuations during the ambipolar diffusion time $\tau_{AD}$ (see
equation [\ref{eq:ndtime}]). If the fluctuations are MHD in origin,
their time scale is given roughly by the Alfv{\'e}n crossing time. 
Sufficiently rapid fluctuations thus impose the constraint 
\be
N_F \equiv {\tau_{AD} \over \tau_{\rm mhd}} = 
\left[ {\gamma C \over 2 (2\pi G)^{1/2}} \right]  
{v_A \over a}  \tau_{AD} \chi^{-1} \gg 1 \, , 
\label{eq:timecon} 
\ee
where $\tau_{AD}$ is the (non-dimensional) time interval required for
diffusion to take place and where $\tau_{\rm mhd}$ is given by
equation [\ref{eq:taumhd}].  The quantity $N_F$ thus represents the
number of times that the distributions are independently sampled
during the course of the diffusion process. The time required for the
field strength to decrease by a factor of e$^1$ $\approx 2.7$ is
usually $\tau_{AD} = 5 - 10$. The ratio $v_A/a$ is typically 3 -- 10,
the length scale $\chi = 1 - 10$, and the dimensionless ratio in
square brackets is about 8. For reasonable values of the parameters,
the left hand side of the above inequality lies in the range $N_F$ =
20 -- 800, a wide range but always comfortably greater than unity. 
Furthermore, we expect the variations from case to case to differ from
the expectation values by relative sizes of order $\sim N_F^{-1/2}$
due to incomplete sampling of the distribution functions during the
diffusion process.  This relative variation is thus expected to be
$\sim 0.04 - 0.2$.

When constraints [\ref{eq:spacecon}] and [\ref{eq:timecon}] are
satisfied, the fluctuations are independent of the spatial derivatives
appearing in the diffusion equation [\ref{eq:diffuse}]. Before time 
averaging the diffusion equation [\ref{eq:diffuse}], we rewrite the 
equation in a specific form to simplify the averaging procedure. 
In particular, we write out the time derivatives, collect the terms 
on the right hand side, and multiply by one factor of $(1+\eta)$ to
obtain the form 
\be
(1 + \xi) {\partial\over\partial\tau} \Bigl( {b \over p} \Bigr) 
+ {b \over p} \Bigl[ {\dot \xi} - {\dot \eta} \, { (1 + \xi) \over 
(1+\eta) } \Bigr] = { (1+\xi)^3 \over (1+\eta)^{3/2} } 
{\partial \over \partial\mu} \left\{ {b^2 \over p^{1/2}} 
{\partial b \over \partial\mu} \right\} \, , 
\ee
where the dots represent time derivatives. With the diffusion equation
written in this form, the time averaging procedure removes all
fluctuations except for an effective diffusion constant on the right
hand side, and the diffusion equation becomes  
\be
{\partial\over\partial\tau} \Bigl( {b \over p} \Bigr) = D
{\partial \over \partial\mu} \left\{ {b^2 \over p^{1/2}} 
{\partial b \over \partial\mu} \right\} \qquad {\rm where} \qquad 
D \equiv \left< {(1+\xi)^3 \over (1+\eta)^{3/2} } \right> \, . 
\label{eq:finaldiff} 
\ee 
Notice that this effective diffusion constant $D$ would be unity in
the absence of fluctuations.

\subsection{Results for particular distributions} 

After time averaging, the diffusion problem is a rescaled version of
the one solved previously. The quasi-static equilibrium state is given
by equation [\ref{eq:kdef}], which contains the factor $K$ due to the
pressure provided by fluctuations. In the diffusion equation, the
effective diffusion constant is larger by the factor $D$. If we
rescale the magnetic field strength according to $\bwig = \sqrt{K} b$,
then the quasi-static equilibrium equation [\ref{eq:kdef}] takes its
standard form. The diffusion equation also takes its standard form if
we rescale the time coordinate according to $\tauwig = (D/K) \tau$. In
other words, the ambipolar diffusion process speeds up by a factor
$\Lambda = D/K$ which is determined by the distribution of the
fluctuations. To make further progress, we thus have to specify the
distributions.

The magnetic field fluctuations $\xi$ and the density fluctuations 
$\eta$ follow normalized distributions, $f(\xi)$ and $g(\eta)$, 
with zero mean, i.e., 
\be
\int f(\xi) d\xi = 1 \, , \quad \int f(\xi) \xi d\xi = 0 \, , 
\quad \int g(\eta) d\eta = 1 \, , \quad {\rm and} \quad 
\int g(\eta) \eta d\eta = 0 \, , 
\ee 
where $f(\xi)d\xi$ is the probability that a fluctuation in the
magnetic field has an amplitude between $\xi$ and $\xi + d\xi$ (the
function $g(\eta)$ is defined similarly for density fluctuations).
For a given choice of distributions $f(\xi)$ and $g(\eta)$, the
expectation value of the diffusion constant $D$, the equilibrium
factor $K$, and the time scale correction factor $\Lambda$ = $D/K$ can
be calculated directly.

For purposes of illustration, we first consider the fluctuations to have 
uniform (flat) distributions with amplitude $A < 1$. In other words, 
\be
f(\xi) = {1 \over 2A} \qquad {\rm for} \quad -A < \xi < A \, , 
\label{eq:fdist}
\ee
and 
\be
g(\eta) = {1 \over 2A} \qquad {\rm for} \quad -A < \eta < A \, . 
\label{eq:gdist}
\ee
We are implicitly assuming that the density fluctuations are
independent of the magnetic field fluctuations. With this particular
choice for the distribution functions, the enhancement factor can be
written in the form  
\be
\Lambda = {6 (1+A^2) \over 3 + A^2} 
{ (1-A)^{-1/2} - (1+A)^{-1/2} \over \ln(1+A) - \ln(1-A) } \, . 
\label{eq:biglambda} 
\ee 
The result is plotted in Figure 1, which shows that the enhancement
factor becomes substantial, $\Lambda \sim 5$, as the fluctuation
amplitude $A$ approaches unity. In this particular treatment, the
enhancement factor has a logarithmic divergence as $A \to 1$.

For comparison, if we ignore density fluctuations and consider only
fluctuations in the magnetic field strength, the enhancement factor
simplifies to the form 
\be
\Lambda = 
{1 + 3 \langle \xi^2 \rangle \over 1 + \langle \xi^2 \rangle} 
= {1 + A^2 \over 1 + A^2/3} \, , 
\label{eq:lambnoeta} 
\ee
where the first equality holds for $\langle \xi^3 \rangle$ = 0 and
where the second equality holds for a uniform distribution of
fluctuations. For the benchmark case $A=1$, the magnetic field
fluctuations are comparable to the total field strength and we find a
modest increase in the ambipolar diffusion rate, i.e., $\Lambda$ =
1.5. In the extreme limit $\langle \xi^2 \rangle \to \infty$, the
enhancement factor $\Lambda \to 3$ (for $\langle \xi^3 \rangle$ =
0). Thus, as expected, fluctuations in the magnetic field strength
alone lead to smaller changes than when the density field has
independent variations.

We can also consider the case of density fluctuations that depend
directly on the magnetic field fluctuations. For example, flux
freezing arguments (Shu 1992) imply that $B \propto \rho^\kappa$,
where $\kappa$ = 2/3 for the usual case of spherical geometry and
$\kappa$ = 1 for a one-dimensional cloud layer. If the fields,
including fluctuations about their mean values, obey such a flux
freezing relation, then we have a correlation of the form $(1+\eta)
\propto (1+\xi)^{1/\kappa}$ and the enhancement factor becomes 
\be 
\Lambda =
{ \langle (1 + \xi)^{3 - 3/2\kappa} \rangle \over 
\langle (1 + \xi)^{2 - 1/\kappa} \rangle } \, .  
\label{eq:lamkappa} 
\ee 
For our reference case of uniform distributions with amplitude $A=1$,
we can evaluate this expression to obtain $\Lambda = (6\kappa-2) 2 /
[(8 \kappa - 3) 2^{1/2\kappa}]$.  For the flux freezing exponent
$\kappa = 1$ appropriate for a one-dimensional layer, for example, we
find $\Lambda \approx 1.13$. 

When the fluctuations in magnetic field strength and density are
correlated, the enhancement factor is generally smaller than in the
absence of correlations. The special case of $\kappa$ = 1/2 is
particularly interesting: As shown by equation [\ref{eq:lamkappa}],
the value $\kappa$ = 1/2 leads to $\Lambda$ = 1, i.e., no net
enhancement of the ambipolar diffusion rate. In other words, for this
particular correlation between the density and magnetic field
fluctuations, the problem reduces to its old (non-fluctuating) form.
This value $\kappa=1/2$ is found in three-dimensional calculations of
magnetic cloud models (Mouschovias 1976) and is consistent with Zeeman
measurements of the magnetic field strength in several molecular
clouds (Crutcher 1999; Basu 2000). However, both the theoretical
calculation and observational relation correspond to the mean field
and not the fluctuations; one interpretation is that the relation 
$B \propto \rho^{1/2}$ is more of an upper envelope than a scaling law
(Zweibel 2001). In any case, the correlation of the magnetic field
fluctuations with the density fluctuations thus needs to be further
specified.

In general, MHD disturbances exhibit a rich variety of different
possible behaviors, including the various kinds of correlations
between density and magnetic field fluctuations. In the linear regime,
for example, pure Alfv{\'e}n waves exhibit magnetic field fluctuations
but have no density variations. Purely acoustic waves have density
variations but no magnetic field fluctuations. When the magnetic field
is perpendicular to the direction of wave propagation, then fluids can
develop magnetosonic waves with correlated magnetic and density
fluctuations; this configuration also supports slow (nonpropagating)
modes that compress the matter (increase density) and displace the
magnetic field.  In real molecular clouds, MHD disturbances are a
complicated and non-linear superposition of many different types of
waves and other motions (one should keep in mind that these different
types of waves propagate at different speeds).  As a result, many
different possible distributions of fluctuations are allowed by the
observations. Without further constraints, the parameter space of
possible fluctuations allows arbitrarily large enhancements in the
ambipolar diffusion rate. This claim is substantiated by equation
[\ref{eq:biglambda}], which shows that $\Lambda$ increases without
bound in the limit $A \to 1$ for one particular choice of
distribution. To obtain bounds on $\Lambda$, we thus need to impose
additional constraints, as discussed below.

\subsection{Application to Molecular Cloud Cores} \label{sec:apply}

We can scale the parameters of this theory to observed molecular
clouds in order to specify the expected regime of our derived results.
As mentioned in the Introduction, molecular clouds have significant
non-thermal contributions to their line-widths (Larson 1981; Myers,
Ladd, \& Fuller 1991) and the implied non-thermal motions are often
interpreted as MHD turbulence.  Indeed, the amplitude of these motions
is consistent with the expected Alfv{\'e}n speed in these regions.

To constrain the allowed range of our fluctuations, we consider the 
observed line-widths to arise from the fluctuating part of the 
Alfv{\'e}n speed in a time-averaged sense. In other words, we assume 
that 
\be
(\Delta v)_{NT}^2 = \left< v_A^2 \right>_{fluc} \, = 
\left< {B^2 \over 4 \pi \rho} {(1+\xi)^2 \over 1 + \eta} \right> - 
\left< {B^2 \over 4 \pi \rho} \right> 
\ee
where we have subtracted off the non-fluctuating part so that the
non-thermal contribution to the line-width vanishes in the absence of
fluctuations. Evaluating this expression in terms of our formalism for
the fluctuations, we find 
\be
(\Delta v)_{NT}^2 = 2 a^2 {b^2 \over p} \bigl[ K - 1 \bigr] = 
2 \alpha_0 a^2 \bigl[ K - 1 \bigr] \, , 
\ee
where $a$ is the sound speed, $K$ measures the amplitude of the
fluctuations as defined by equation [\ref{eq:kdef}], and $\alpha_0$ is
the initial ratio of magnetic to thermal pressure.  To evaluate the
ratio $b^2/p$, we have used our equilibrium state with no fluctuations
as a reference state. Since the observed line-widths are comparable to
the expected Alfv{\'e}n speeds, we expect $(\Delta v)_{NT}^2 \approx 2
\alpha_0 a^2$ and thus we expect the quantity $(K-1)$ to be close to
unity.  In other words, scaling our formulation to observed molecular 
clouds implies that $K \approx 2$. 

We need to apply this result to constrain the possible values of
$\Lambda$. For fluctuations obeying uniform distributions, the
parameter $K$ = 2 for an amplitude $A \approx 0.886$; the
corresponding value of the enhancement factor is $\Lambda \approx
2.26$.  For the special case of no density fluctuations, the
enhancement factor is given by equation [\ref{eq:lambnoeta}]; the
constraint $K=2$ implies that $\langle \xi^2 \rangle$ = 1 and hence
$\Lambda$ = 2 (in the symmetric limit where $\langle \xi^3 \rangle$ =
0). In general, the inclusion of density fluctuations makes the
enhancement factor larger. For typical star forming regions, our net
result is that fluctuations increase the ambipolar diffusion rate by a
factor of 2.

\subsection{Comparison to Numerical Results} 

We now compare our analytic results with numerical simulations.
Instead of working exclusively with the expectation values ($D$ and
$K$), we can use the same distributions for the fluctuations $\xi$ and
$\eta$ and then sample the distributions as we numerically integrate
the equations of motion. Specifically, we use a time-averaged
quasi-static equilibrium state (equation [\ref{eq:kdef}]) and
numerically integrate the diffusion equation [\ref{eq:finaldiff}] by
sampling the distribution functions for the fluctuations to specify
the effective diffusion constant.  Our basic numerical scheme is
similar to that described in Shu (1983), although the time resolution
must be significantly higher (and highly variable) to properly include
the wide range of values for the effective diffusion constant. Each
numerical simulation results in one particular realization of the time
evolution. If we perform enough realizations of the problem, the
average time scale should be the same as that calculated from the
expectation values above.

To numerically follow the evolution of the cloud, we need to specify
the starting condition. Following previous authors, we adopt the
following modified standard family of initial states 
\be
p = {1\over 1 +\alpha_0} \, {\rm sech}^2 
\left({y\over 1+\alpha_0}\right) = {1\over 1+\alpha_0} 
\left(1 - \mu^2\right)\, ,
\ee
\be
b = \left[{\alpha_0 \over (1 +\alpha_0)K}\right]^{1/2} 
\, {\rm sech} \left({y\over 1+\alpha_0}\right)
= \left[{\alpha_0\over (1+\alpha_0)K}\right]^{1/2} 
\left(1 - \mu^2\right)^{1/2}\, ,
\ee
\be
\mu = \tanh \left({ y \over 1 + \alpha_0}\right)\, ,
\ee
where the ratio $\alpha_0/K$ represents the initial 
magnetic to thermal pressure ratio, i.e., 
\be
{\alpha_0 \over K} = {b^2\over p} = {B^2 / 8 \pi \over a^2 \rho}\, .
\ee
Through the action of ambipolar diffusion, the fluid evolves toward a
final state described by the above equations in the limit 
$\alpha_0 \to 0$. 

In addition to recovering the expectation values for the diffusion
time, the numerical treatment also finds the deviations about the mean
value (the expectation value). These deviations depend on how often we
sample the distributions to evaluate the effective diffusion constant.
If we sample the distributions often enough, say at every numerical
time step, then the numerically calculated time scale converges to the
expectation values found earlier.  As outlined in \S 3.3, however, we
expect the fluctuations to vary on the MHD crossing time scale
(equation [\ref{eq:taumhd}]). In dimensionless time units, we thus
expect the diffusion constant to take on independent new values for
time intervals longer than $\tau_X \approx 0.1 a/v_A \sim 0.02$. For
the sake of definiteness, we take $\tau_X$ = 0.02 for all of the
simulations presented here. As we show below, however, the value 
of $\tau_X$ determines the width of the distribution of possible 
time scales; we can estimate this width analytically and then scale 
our results to other choices of $\tau_X$. 
 
We compare our numerical results with the analytical formulation in
Figure 2. We have used uniform distributions of the fluctuations with
amplitude $A$ (see equations [\ref{eq:fdist} -- \ref{eq:gdist}]) and
set $\alpha_0 = 10$ so that the magnetic pressure is ten times larger
than the thermal pressure at the start of the calculation.  In these
simulations, the diffusion constant $D_j$ = $(1+\xi)^3/(1+\eta)^{3/2}$
is changed every time interval $\tau_X = 0.02$ by sampling the
distributions of $\xi$ and $\eta$. The resulting time scale $\tau_{\rm
e}$ for the magnetic field strength to decrease by one e-folding is
plotted versus the fluctuation amplitude in Figure 2. The solid curve
shows the expectation value for the time scale $\tau_{\rm e} =
\tau_{{\rm e}0} / \Lambda$, where $\tau_{{\rm e}0}$ is the time scale
in the absence of fluctuations.  The symbols show different
realizations found by numerically integrating the problem. For this
set of simulations, we have selected the amplitude $A$ randomly as
well. As expected, the numerical integrations agree with the analytic
predictions of \S 3.3. In particular, the mean value of the
distribution of time scales from the numerical experiments closely
follows the expectation value.

The width of the distribution (for a given amplitude $A$) increases
with the amplitude of the fluctuations. As as result, fluctuations not
only force ambipolar diffusion to take place more rapidly, on average,
but they also allow the process to sample a range of time scales. This
range of possible time behaviors, as illustrated in Figure 2, is
determined (in part) by the time scale $\tau_X$ that specifies how
often the effective diffusion constant takes on different values. In
the limit $\tau_X \to 0$, the ambipolar diffusion time scale
approaches the expectation value and the width of the time scale
distribution shrinks to zero. In the opposite limit, the effective
diffusion constant takes on a single (but randomly selected) value for
the whole evolutionary time. In dimensionless units, the e-folding
time for ambipolar diffusion is $\tau_{\rm e}$ = 5 -- 10 in this
regime of parameter space. With $\tau_X$ = 0.02, near the low end of
the range estimated from MHD considerations, the effective diffusion
constant takes on $N$ = 250 -- 500 different values during an
evolutionary run.

We can quantify the width of the distribution of time scales for a
given distribution of fluctuations and a given amplitude. For purposes
of illustration, we use uniform distributions of fluctuations with
amplitude $A$ = 0.886, which corresponds to the case in which the
fluctuations are large enough to account for the observed non-thermal
line widths in molecular clouds (see \S 3.5). Figure 3 shows the
distribution of ambipolar diffusion time scales, calculated both
numerically and analytically. The numerical results were obtained by
running the aforementioned code 10,000 times to build up the
distribution shown by the solid curve in Figure 3. For comparison, the
dashed curve shows a gaussian distribution whose width is calculated
as described below. The solid spike at $\tau \approx 12.5$ represents
the delta-function distribution that applies in the absence of
fluctuations.

The distribution of time scales $\tau$ can be estimated analytically.
The peak of the distribution (denoted here as $\langle \tau \rangle$)
is determined by the enhancement factor $\Lambda$ calculated in \S
3.4. For the case shown in Figure 3, $\Lambda = 2.25$ and hence
$\langle \tau \rangle$ = 5.54. We can also find the width of the
distribution.  For a given realization of the problem, the time scale
$\tau$ is determined by the effective value $\dwig$ of the diffusion 
constant, i.e., 
\be
\dwig = D + {D \over N_F} \sum_{j=1}^{N_F} 
\bigl[ {D_j \over D} - 1 \bigr] \, , 
\ee 
where $N_F$ is the number of independent samplings of the diffusion
constant (equation [\ref{eq:timecon}]), $D$ is the expectation value,
and the $D_j$ are the particular choices taken during a given run.
(Notice that in the limit $N_F \to \infty$, we must have $\dwig \to
D$.)  For simplicity, we consider $N_F = \langle \tau \rangle / \tau_X$ 
to have a fixed value, although the exact value also varies from case 
to case (this complication is a higher order effect). The case-to-case 
variation $\Delta D$ = $\dwig - D$ in the diffusion constant is thus
proportional to the sum of a large number (here, $N_F \sim 300$) of
random variables $\zeta_j$ = $D_j/ D - 1$. These variables $\zeta_j$
are constructed to have zero mean and follow a distribution given by
the convolution of $f(\xi)$ and $g(\eta)$; we denote the variance of
this $\zeta_j$ distribution as $\sigwig$.  Because of the central
limit theorem, the distribution of the composite variable $\Delta D$
takes a gaussian form. Similarly, distribution of time scales takes a
gaussian form and its width is given by
\be 
\langle \sigma \rangle = \sigwig \, 
\sqrt{\tau_X \langle \tau \rangle} \, . 
\label{eq:width} 
\ee
We can compare this formula with our numerical results for the
particular case shown in Figure 3 (uniform fluctuations with $A$ =
0.886).  The dashed curve shows a gaussian distribution with this
predicted width; the distribution is in good agreement with the
histogram of results (the solid curve) from the numerical simulations.
More precisely, we find that the expectation value for the ambipolar
diffusion time scale $\tau_{\rm e} = \tau_{{\rm e}0} / \Lambda$
$\approx 5.54$ is in good agreement with that found numerically, i.e., 
$\langle \tau \rangle$ $\approx$ 5.62. The chosen distributions of
fluctuations show that $\sigwig \approx 2.51$ and hence equation
[\ref{eq:width}] implies that the distribution of time scales should
have width $\langle \sigma \rangle \approx$ 0.835. The numerical
simulations imply almost the same value ($\langle \sigma \rangle
\approx$ 0.834). For all quantities that can be compared, the 
numerical results and the analytic predictions agree to within 
about one percent. 

As shown by equation [\ref{eq:width}], the width of the time scale
distribution depends on the time scale $\tau_X$ over which the
fluctuations occur, or, equivalently, the number $N_F$ of independent
fluctuation samples. The value chosen for our numerical simulations,
$\tau_X = 0.02$, is near the low end of the expected range. For
comparison, the dotted curve in Figure 3 also shows the distribution
for a $\tau_X = 0.055$, which corresponds to $N_F = 100$. Notice that
as the fluctuation time scale $\tau_X$ grows even larger (and $N_F$
decreases), the distribution of time scales grows wider and eventually
departs from a gaussian form.

\section{Discussion} \label{sec:discuss}

In this paper, we have explored how fluctuations in the background
fields affect the rate of ambipolar diffusion. These fluctuations
force the magnetic field strength and the density to sample a
distribution of values, rather than take on a single value at a given
point in space and time. We have used a one-dimensional molecular
cloud layer as a test problem to study the effects of such
fluctuations.

The first principal result of this paper is that the time scale for
ambipolar diffusion is altered by these fluctuations. In particular,
fluctuations drive ambipolar diffusion to take place more rapidly,
with the time scale shorter by a factor $\Lambda \sim 1 - 10$. For
typical conditions in molecular clouds cores, the enhancement factor
is near the lower end of this range, $\Lambda \sim 2 - 3$, but much
larger enhancements remain possible.  The ambipolar diffusion time
scale depends on the distribution of fluctuations. For the case of
uniform distributions, for example, the ambipolar diffusion time scale
varies with the amplitude $A$ as shown in Figure 1. In general, the
time scale also depends on the shape of the distributions.

For a given distribution of fluctuations, the ambipolar diffusion time
scale also varies from realization to realization (see Figure 2).
Descriptions of the ambipolar diffusion process thus face an
interesting complication, which is our second principal result: The
time scale for loss of magnetic support takes on a {\it distribution}
of values instead of a single value.  Consider two identical molecular
cloud regions and suppose they are laced with fluctuations following a
given distribution. Because the two regions experience incomplete (and
different) samplings of the fluctuation distributions as their
magnetic fields diffuse outwards, they will not exhibit the same
diffusion time.  This feature is more general than its manifestation
in this particular test problem.  When a physical system contains an
effectively random element -- in this context through chaos and
turbulence -- the outcomes must be described in terms of a probability
distribution. For our test problem (ambipolar diffusion in a cloud
layer), the single value of the e-folding time $\tau_{\rm e}$ is
replaced by a distribution of values (see Figure 3). Furthermore, the
distribution of possible time scales approaches a gaussian form; the
most likely value for the time scale is shorter than the case without
fluctuations by the enhancement factor $\Lambda = D/K$ (see equations
[\ref{eq:kdef}, \ref{eq:finaldiff}]) and the width of the distribution
is given by equation [\ref{eq:width}].

This effect on the ambipolar diffusion time scale has important
implications for star formation in molecular clouds. These clouds
appear to be supported by magnetic fields and the observed magnetic
field strengths are commensurate with this view. However, statistics
of molecular cloud cores (with and without young stellar objects)
argues that the (uncorrected) time scale for ambipolar diffusion may
be too long to account for the observations. This work shows that
magnetic fields can diffuse more rapidly than previous estimates
suggest. This speed-up, along with any other enhancements (e.g.,
Ciolek \& Basu 2001), can help account for the observed statistics of
molecular cloud cores.  Another complicating issue arises: Because the
ambipolar diffusion time scale takes on a distribution of values, and
this distribution can be rather wide if the fluctuations change on
long time scales $\tau_X$, some core regions will experience much
faster diffusion rates than others even if they have ``the same''
starting conditions.  In this regime of diffusion activity, the cores
that actually form stars are those which evolve on the ``fast'' side
of the distribution, whereas the cores that happen to live on the
``slow'' side of the distribution will fail to form new stars.

This preliminary treatment of fluctuations, including their effects on
ambipolar diffusion and star formation, remains incomplete in several
respects. In this paper, we have separated the calculation of the
diffusion process from the determination of the fluctuations. In
particular, we have assumed {\it a priori} forms for the fluctuations
to study their implications. In a complete treatment, one should
calculate the fluctuations and their effects in a self-consistent
manner. In addition, we have focused on long wavelength fluctuations
and have not considered spatial gradients in the fluctuating part of
the fields. Magnetic turbulence cascades down to small scales,
however, so it is possible that fields fluctuate at length scales
smaller than our MHD condition. This complication should also be
considered in future work. Our present treatment is limited to
one-dimensional slab models so that magnetic tension is not included;
two-dimensional simulations should be done in the future.  Another
classical problem is the heating of molecular cloud regions by
ambipolar diffusion; the effects of fluctuations on this mechanism
should be considered. Finally, the act of star formation provides a
source of new turbulence, which drives new fluctuations and can affect
the ambipolar diffusion rates of neighboring cores; this feedback
effect should also be studied.  In any case, fluctuations in both the
magnetic and density fields introduce an effectively random element
into the ambipolar diffusion process, and thereby provide a rich class
of new behavior for further study.

\bigskip 
%\newpage 
\centerline{\bf Acknowledgements} 

We would like to thank Charlie Doering, Phil Myers, Steve Shore, and
Frank Shu for useful discussions. We also thank the referee -- Glenn 
Ciolek -- for comments that improved the paper.  MF is supported by
the Hauck Foundation through Xavier University. FCA is supported by
NASA through a grant from the Origins of Solar Systems Program and by
Univ. Michigan through the Michigan Center for Theoretical Physics.

\bigskip
\centerline{\bf APPENDIX A: Cylindrical Geometry}
\medskip

In this Appendix, we consider the effects of fluctuations on ambipolar
diffusion in a molecular cloud filament. We consider only the simplest
case of magnetic field lines that are aligned with the axis of the
filament and depend only on the radial coordinate $r$, i.e., we have 
$$
{\bf B} = B(r) \, {\hat z} \, . 
\eqno({\rm A}1)$$ 
With this basic configuration, the equations of motion take the form 
$$
{\partial \rho \over \partial t} + {1 \over r}
{\partial \over \partial r} (r \rho u) = 0 \, ,
\eqno({\rm A}2)$$
$$
{\partial u\over\partial t} + u {\partial u\over
\partial r} = g - {1\over\rho}{\partial\over\partial r}
\left(P+{B^2\over 8\pi}\right)\, ,
\eqno({\rm A}3)$$
$$
{1\over r}{\partial\over\partial r}(rg) = -4\pi G\rho\,,
\eqno({\rm A}4)$$
and finally
$$
{\partial B \over \partial t} + {1 \over r} 
{\partial \over \partial r} (r B u) =  {1 \over 4 \pi \gamma C} 
{1 \over r} {\partial \over \partial r}  \Bigl\{ {r \over \rho^{3/2}} 
B^2 {\partial B \over \partial r}  \Bigr\} \, , 
\eqno({\rm A}5)$$
where we have defined $u$ to be the radial (and only nonvanishing) 
component of the velocity, and we have made use of the relationship
defined by equation [\ref{eq:rhoion}]. In practice, these filaments 
will be subject to clumping instabilities in both the linear (Gehman, 
Adams, \& Watkins 1996) and nonlinear regimes (Adams, Fatuzzo, \& 
Watkins 1994); in this derivation, however, we neglect this issue 
and focus on the effects of the cylindrical geometry. 

Next, we introduce the fluctuations through the ansatz given by
equation [\ref{eq:ans}] and rewrite the problem in terms of a
Lagrangian description of the dynamics (e.g., see \S 2).  For the
cylindrical geometry considered here, the relevant Lagrangian
coordinate is the mass per unit length $\sigma$ along the filament
within a radius $r$ of the central axis, i.e., 
$$
\sigma \equiv \int_0^r \, \rho(r', t) \, r' \, dr' \, . 
\eqno({\rm A}6)$$
Notice that the variable $\sigma$ differs from the true mass 
per unit length by a factor of $2 \pi$ which has been omitted 
for simplicity.  The original problem in the variables $(r, t)$ 
is now transformed to one in new variables $(\sigma, t)$ and the 
derivatives transform according to 
$$
{\partial \over \partial t} + u 
{\partial \over \partial r} \to \, 
{\partial \over \partial t}\Bigg|_\sigma \, , 
\eqno({\rm A}7)$$
$$
{\partial \over \partial r} \to \, 
r \rho {\partial \over \partial \sigma}\Bigg|_t \, , 
\eqno({\rm A}8)$$
$$
u \to {\partial r \over \partial t}\Bigg|_\sigma \, , 
\label{eq:concyl}
\eqno({\rm A}9)$$

With this transformation, the equation of continuity becomes 
$$
{\partial r \over \partial \sigma} = {1 \over r \rho} \, . 
\eqno({\rm A}10)$$
For an isothermal equation of state, the force equation can be
written in the form
$$
- {1 \over r} \, {\partial^2 r \over \partial t^2} =  
{4\pi G\over r^2} \int_0^r \rho (1 + \eta) r' dr'
+ {a^2\over 1+\eta} {\partial \over \partial \sigma} 
[\rho(1+\eta)] + {1 \over 1+\eta} 
{\partial  \over \partial \sigma} \left[{B^2
(1+\xi)^2\over 8\pi}\right]\, , 
\eqno({\rm A}11)$$
and the nonlinear diffusion equation for the magnetic 
field becomes 
$$ 
{\partial \over \partial t} \left[{B(1+\xi)\over \rho(1+\eta)}
\right]=  {1\over 1+\eta}{\partial \over \partial \sigma}  
\Bigl\{ {r^2 B^2 (1+\xi)^2\over 4\pi \gamma C \rho^{1/2}
(1+\eta)^{3/2}}{\partial  \over \partial \sigma}  [B(1+\xi)]
\Bigr\} \, . 
\label{eq:difcyl}
\eqno({\rm A}12)$$ 
Equations [A10 -- A12] exhibit exactly the same form as their slab
counterparts (equations [\ref{eq:continuity} -- \ref{eq:diffuse0}]). 
As such, the effects of fluctuations on ambipolar diffusion in a
cylindrical filament will scale exactly as for the slab geometry
considered in the main text.

%\newpage 
\bigskip
\centerline{\bf APPENDIX B: Short Wavelength Fluctuations} 
\medskip

We consider the effects of short wavelength fluctuations in this
Appendix.  As noted in the main body of the text, this analysis is
complicated by the fact that the solutions to stochastic differential
equations depend on the manner in which various limits are taken
(Doering 1990). The formulation presented below thus represents one
possible approach to the general problem.

The most likely source of short wavelength fluctuations is MHD
turbulence, which is present in most regions of molecular clouds. The
MHD condition, already built into the ambipolar diffusion equations,
requires that the neutral fluid remain coupled to the ions and to the
magnetic field.  Physically, this condition is met if the ion-neutral
collision frequency $f_{in} = \gamma \rho_i$ exceeds the frequency
associated with the MHD turbulence. The latter frequency can be
approximated by $f_{mhd} \approx v_A / \lambda$, where $v_A = B / (4
\pi \rho)^{1/2}$ is the Alfv\'en wave speed and $\lambda$ is the
length scale of the fluctuations.  As a result, the coupling condition
requires that 
$$
\chi > 0.09 {b\over p}\, , 
\eqno({\rm B}1)
$$
where $\chi$ is the dimensionless turbulence length scale
as defined by equation [\ref{eq:chidef}].

If we assume that the fluctuations are both spatially and temporally
symmetric (which means that $\xi$ and $\eta$ are not correlated with
their first order derivatives), then the following relations hold: 
$$
\left< F(\xi, \eta) {\partial\eta\over\partial\mu} \right>
\approx 0 \qquad {\rm and} \qquad \left< F(\xi, \eta)
{\partial \xi \over\partial\mu} \right> \approx 0 \, ,
\eqno({\rm B}2) 
\label{eq:fslope}
$$
and 
$$
\left< F(\xi, \eta) {\partial\eta\over\partial\tau} \right>
\approx 0 \qquad {\rm and} \qquad \left< F(\xi, \eta)
{\partial \xi \over\partial\tau} \right> \approx 0 \, ,
\eqno({\rm B}3)
$$
for all well-behaved functions $F(\xi, \eta)$. Under these conditions,
the quasi-equilibrium state described in \S 3.2 remains valid.

For short wavelength fluctuations, expanding the diffusion equation
using the same approach as presented in \S 3.3 yields the form 
$$ 
(1 + \xi) {\partial\over\partial\tau} \Bigl( {b \over p} \Bigr) 
+ {b \over p} \Bigl[ {\dot \xi} - {\dot \eta} \, { (1 + \xi) \over 
(1+\eta) } \Bigr] = { (1+\xi)^3 \over (1+\eta)^{3/2} } 
{\partial \over \partial\mu} \left\{ {b^2 \over p^{1/2}} 
{\partial b \over \partial\mu} \right\} 
+ \left\{ {\partial\over\partial\mu}
\left[{(1+\xi)^3\over (1+\eta)^{3/2}}\right] \right\}
\left( {b^2 \over p^{1/2}} {\partial b\over\partial\mu}\right)
$$
$$ 
+{(1+\xi)^2 \over (1+\eta)^{3/2}} \left\{ {\partial\over\partial\mu} 
(1+\xi) \right\} \left\{ {\partial\over\partial\mu}
\left[{b^3\over p^{1/2}}\right]\right\}
+{b^3 \over p^{1/2}} {\partial\over\partial\mu}
\left\{{(1+\xi)^2\over (1+\eta)^{3/2}} 
{\partial\over\partial\mu} (1+\xi)\right\} \, . 
\eqno({\rm B}4) 
\label{eq:bigdiff}
$$
We note that the second and third terms on the right hand side vanish
when they are time averaged because they take the form given by
equation [\ref{eq:fslope}]. With this simplification, the
time-averaged diffusion equation reduces to the form 
$$
{\partial\over\partial\tau}
\left({b\over p}\right) = 
D {\partial\over\partial\mu}\left( {b^2\over p^{1/2}}
{\partial b\over\partial\mu}\right) + G {b^3\over p^{1/2}}\,,
\eqno({\rm B}5) 
\label{eq:fdiffsw} 
$$
where $D$ is defined in equation [\ref{eq:finaldiff}] and
$$
G = \left<{\partial\over\partial\mu}
\left[{(1+\xi)^2\over (1+\eta)^{3/2}} {\partial\over\partial\mu}
(1+\xi)\right]\right>\,.
\eqno({\rm B}6) 
$$
For the case in which the fluctuations $\xi$ and $\eta$ are 
uncorrelated, the parameter $G$ simplifies to the form 
$$
G = {1\over 3}\left<{1\over (1+\eta)^{3/2}}\right>
\left<{\partial^2 \over \partial\mu^2} (1+\xi)^3\right> \, .
\eqno({\rm B}7) 
$$
Similarly, for the case of perfectly correlated fluctuations, 
$G$ simplifies to the form 
$$
G = {2\over 3} \left<{\partial^2 \over \partial\mu^2}
(1+\xi)^{3/2}\right> \, .
\eqno({\rm B}8) 
$$ 

The relative size of the two terms on the right hand side of equation
[B5] ultimately determines the behavior of the magnetic field
diffusion. Since the fluctuations $\xi$ and $\eta$ vary on a length
scale $\chi \ll 1$, whereas $b$ and $p$ vary on a lengthscale $1 +
\alpha_0 \approx K v_A^2 / a^2 \gg 1$, a simple scaling analysis
naively suggests that the second term (with coefficient $G$) would
dominate over the first (with coefficient $D$). Upon closer
inspection, however, we see that the derivatives of the fluctuations
tend to cancel out, so that the relative sizes of $D$ and $G$ depend
on the form of the fluctuations. In any case, however, this treatment
does not yield an expression that can be scaled to the previous
solutions with no fluctuations.

{} 

%\vskip 0.5truein 
\newpage 
\centerline{\bf FIGURE CAPTIONS} 
\bigskip 

\noindent 
Figure 1. Enhancement factor as a function of fluctuation amplitude.
The solid curve shows the factor $\Lambda$ by which the ambipolar
diffusion process is sped up by fluctuations, as a function of their
amplitude $A$ for uniform distributions. The dashed curves shows the
enhancement factor for fluctuations in the magnetic field only,
whereas the dotted curve shows the result for density fluctuations. 
Notice that fluctuations in the magnetic field produce a greater
effect for relatively small amplitudes, but the density fluctuations
are more important for larger amplitudes.

\noindent 
Figure 2. Comparison of numerical and analytic results. The time scale
$\tau_{\rm e}$ is the e-folding time for the magnetic field strength
to decrease. The time $\tau_{\rm e}$ is plotted here as a function of
fluctuation amplitude $A$ for uniform distributions of magnetic and
density fluctuations. The solid curve shows the expectation value of
the e-folding time scale as calculated analytically in \S 3.3. The
symbols show results from numerical integrations of the same problem
for different samplings of the distributions (see \S 3.5), where we 
have used $\tau_X$ = 0.02 as the time scale over which the diffusion 
constant changes (see text). 

\noindent 
Figure 3. Distribution of ambipolar diffusion times for a cloud layer
with uniform fluctuations of amplitude $A$ = 0.886; this level of
fluctuations is consistent with the non-thermal line widths observed
in star forming regions. The solid curve (histogram) shows the result
of 10,000 numerical simulations with different realizations of the
fluctuations. The dashed curve shows the analytic prediction for the
time scale distribution -- a gaussian with a peak value given by
the expectation value and with a width predicted by application of 
the central limit theorem. The dotted curve depicts a wider gaussian
distribution that applies for longer fluctuation time scales (here,
$N_F$ = 100 or $\tau_X \approx 0.055$). In the absence of
fluctuations, the cloud maintains a single value for its ambipolar
diffusion time, as shown by the delta-function spike at $\tau_{\rm e}
\approx 12.5$.


\newpage 
\begin{thebibliography}{} 

\bibitem[Adams et al. 1994]{AFW94} 
Adams, F. C., Fatuzzo, M., \& Watkins, R. 1994, {\sl ApJ}, {\bf 426}, 629

\bibitem[Aronmax 1975]{am75} 
Arons, J., \& Max, C. E. 1975, {\sl ApJ}, {\bf 196}, L77  

\bibitem[Basu 1997]{basu97}
Basu, S. 1997, {\sl ApJ}, {\bf 485}, 240 

\bibitem[Basu 1998]{basu98}
Basu, S. 1998, {\sl ApJ}, {\bf 509}, 229 

\bibitem[Basu 2000]{bas00}
Basu, S. 2000, {\sl ApJ}, {\bf 540}, L103

\bibitem[Basu 1994]{bas94}
Basu, S., \& Mouschovias, T. Ch. 1994, {\sl ApJ}, {\bf 432}, 720

\bibitem[Caselli et al. 2001]{cas01} 
Caselli, P. et al. 2001, submitted to {\sl ApJ}, astro-ph/0109023  

\bibitem[Ciolek and Basu 2000]{ciobas00}
Ciolek, G. E., \& Basu, S. 2000, {\sl ApJ}, {\bf 529}, 925

%\bibitem[Ciolek and Basu 2000b]{ciobas00b}
%Ciolek, G. E., \& Basu, S. 2000b, in From Light Darkness to Light, 
%ASP Conf. Ser. Vol. 243, p. 79 

\bibitem[Ciolek \& Basu 2001]{cb2001} 
Ciolek, G. E., \& Basu, S. 2001, {\sl ApJ}, {\bf 547}, 272 

\bibitem[Ciolek \& Mouse 1994]{cm94} 
Ciolek, G. E., \& Mouschovias, T. Ch. 1994, {\sl ApJ}, {\bf 425}, 142 

\bibitem[Ciolek \& Mouse 1994]{cm95} 
Ciolek, G. E., \& Mouschovias, T. Ch. 1995, {\sl ApJ}, {\bf 454}, 194 

\bibitem[Ciolek \& Mouse 1998]{cm98} 
Ciolek, G. E., \& Mouschovias, T. Ch. 1998, {\sl ApJ}, {\bf 504}, 280 

\bibitem[Crutcher 1998]{cr98}
Crutcher, R. M. 1998, in {\sl Interstellar Turbulence}, eds. J. Franco and 
J. Carraminana (Cambridge: Cambridge Univ. Press), pp. 213 

\bibitem[Crutcher 1999]{cr99}
Crutcher, R. M. 1999, {\sl ApJ}, {\bf 520}, 706 

\bibitem[Doering 1990]{dor90} 
Doering, C. R. 1990, in {\sl Complex Systems Summer School}, ed. 
L. Nadel \& D. Stein (Addison Wesley), pp. 3 -- 51 

\bibitem[Doering et al. 1994]{dhr94} 
Doering, C. R., Horsthemke, W., \& Riordan, J. 1994, 
{\sl Phys. Rev. Lett.}, {\bf 72}, 2984 

\bibitem[Fatuzzo \& Adams 1993]{fa93}
Fatuzzo, M., \& Adams, F. C. 1993, {\sl ApJ}, {\bf 412}, 146 

\bibitem[gammie1996]{go96}
Gammie, C. F., \& Ostriker, E. C. 1996, {\sl ApJ}, {\bf 466}, 814 

\bibitem[Gehman et al. 1996]{GAW96} 
Gehman, C. S., Adams, F. C., \& Watkins, R. 1996, {\sl ApJ}, {\bf 472}, 673 

\bibitem[Greaves 1999]{gh99}
Greaves, J. S., \& Holland, W. S. 1999, {\sl MNRAS}, {\bf 302}, L45 

\bibitem[Jijina et al. 1999]{JMA99}
Jijina, J., Myers, P. C., \& Adams, F. C. 1999, {\sl ApJ Suppl.}, 
{\bf 125}, 161 

\bibitem[Jokipii \& Parker 1969]{jp69}
Jokipii, J. R., \& Parker, E. N. 1969, {\sl ApJ}, {\bf 155}, 777 

\bibitem[Kim 1997]{kim97} 
Kim, E.-J. 1997, {\sl ApJ}, {\bf 477}, 183

\bibitem[larson 1985]{larson85}
Larson, R. B. 1981, {\sl MNRAS}, {\bf 194}, 809

\bibitem[Lizano \& Shu 1989]{ls89} 
Lizano, S., \& Shu, F. H. 1989, {\sl ApJ}, {\bf 432}, 834 

\bibitem[mz]{mz95}
McKee, C. F., \& Zweibel, E. G. 1995, {\sl ApJ}, {\bf 440}, 686  

\bibitem[Mouschovias 1976]{mous76}
Mouschovias, T. Ch. 1976, {\sl ApJ}, {\bf 207}, 141

\bibitem[Mouschovias 1982]{mous82}
Mouschovias, T. Ch. 1982, {\sl ApJ}, {\bf 252}, 193  

\bibitem[Mouschovias 1983]{mous83}
Mouschovias, T. Ch. 1983, in IAU Symp. 102, Solar and Stellar Magnetic 
Fields, ed. J. Stenflo, p. 479

%\bibitem[Mouschovias 1985]{mous85}
%Mouschovias, T. Ch. et al. 1985, {\sl ApJ}, {\bf 291}, 772

%\bibitem[Mouschovias 1999]{mous99}
%Mouschovias, T. Ch., \& Ciolek, G. E. 1999, in The Origin of Stars and
%Planetary Systems, ed. C. Lada and N. Kylafis, p. 305

\bibitem[Mouschovias \& Morton 1991]{mous91}
Mouschovias, T. Ch., \& Morton, S. 1991, {\sl ApJ}, {\bf 371}, 296 

\bibitem[Mouschovias 1995]{mous95}
Mouschovias, T. Ch., \& Psaltis, D. 1995, {\sl ApJ}, {\bf 444}, L105 

\bibitem[mg]{mg99}
Myers, P. C., \& Gammie, C. F. 1999, {\sl ApJ}, {\bf 522}, 141

\bibitem[Myers \& Goodman 1988]{mg88}
Myers, P. C., \& Goodman, A. A. 1988, {\sl ApJ}, {\bf 326}, L27  

\bibitem[mlf]{mlf91}
Myers, P. C., \& Ladd, E. F., \& Fuller, G. A. 1991, {\sl ApJ}, 
{\bf 372}, 95

\bibitem[ml]{ml99}
Myers, P. C., \& Lazarian, A. 1998, {\sl ApJ}, {\bf 507}, 157

\bibitem[Nakano 1984]{nk84}
Nakano, T. 1984, {\sl Fund. Cosmic Phys.}, {\bf 9}, 139

\bibitem[Press et al. 1986]{numrep86}
Press, W. H., Flannery, B. P., Teukolsky, S. A., \& Vetterling, W. T.
1986, {\sl Numerical Recipes: The art of scientific computing}
(Cambridge: Cambridge Univ. Press)

\bibitem[Shore \& Adelman]{sa76} 
Shore, S. N., \& Adelman, S. J. 1976, {\sl ApJ}, {\bf 209}, 816 

\bibitem[Shu 1983]{sh83}
Shu, F. H. 1983, {\sl ApJ}, {\bf 273}, 202 

\bibitem[Shu 1992]{sh92}
Shu, F. H. 1992, {\sl Gas Dynamics} (Mill Valley: Univ. Science Books) 

\bibitem[Shu et al. 1987]{sal87}
Shu, F. H., Adams, F. C., \& Lizano, S. 1987, {\sl A R A \& A}, 
{\bf 25}, 23

\bibitem[Soong 1973]{sog73}
Soong, T. T. 1973, {\sl Random Differential Equations in Science and
Engineering} (New York: Academic Press)

\bibitem[Srin \& Vas]{sv71} 
Srinivasan, S. K., \& Vasudevan R. 1971, {\sl Introduction to Random
Differential Equations and their Applications} (New York: Elsevier)

\bibitem[Zweibel 1988]{zw88}
Zweibel, E. G. 1988, {\sl ApJ}, {\bf 329}, 384 

\bibitem[Zweibel 2001]{zw01}
Zweibel, E. G. 2001, submitted to {\sl ApJ}, astro-ph/0107462 

\end{thebibliography}
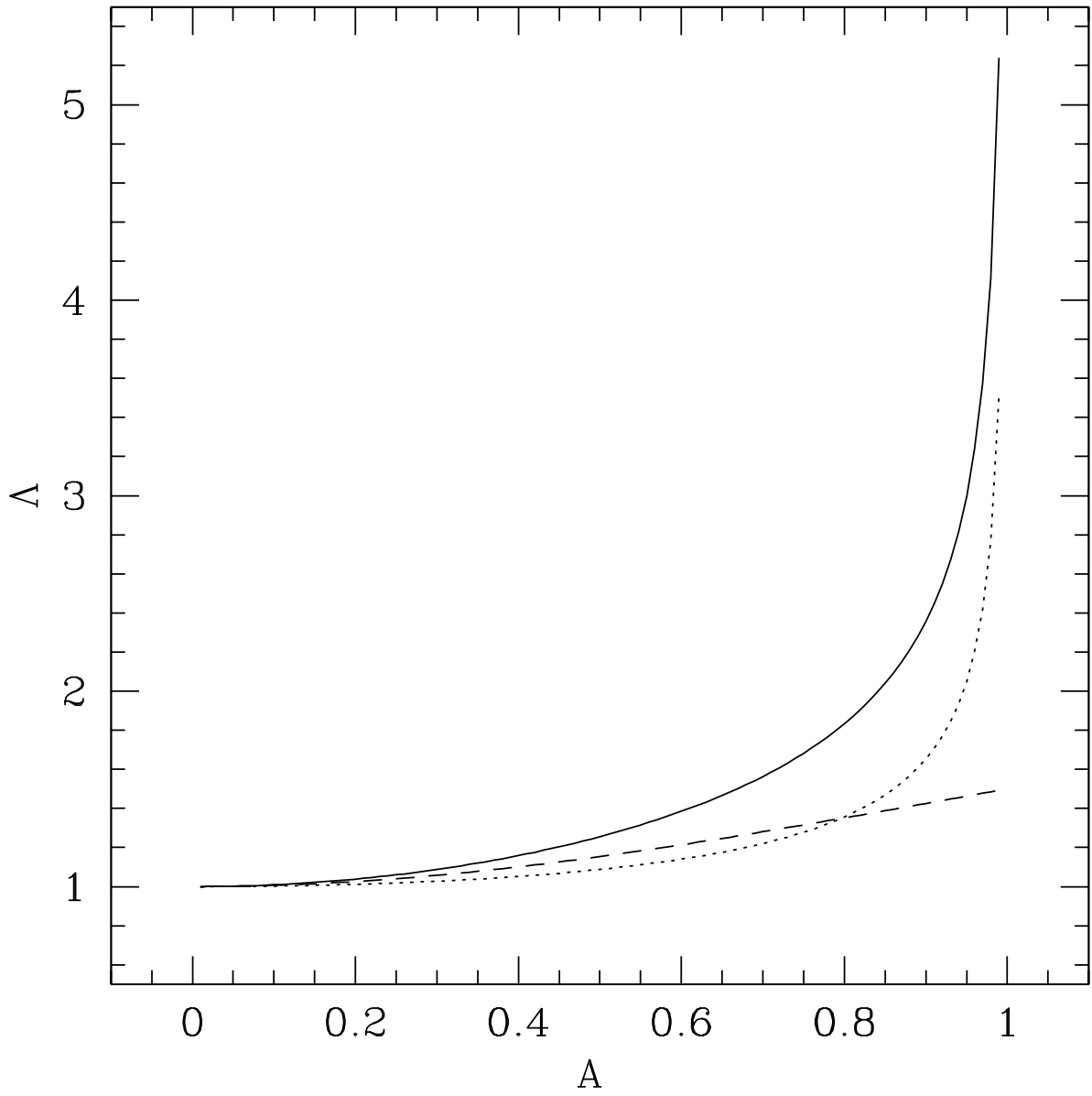
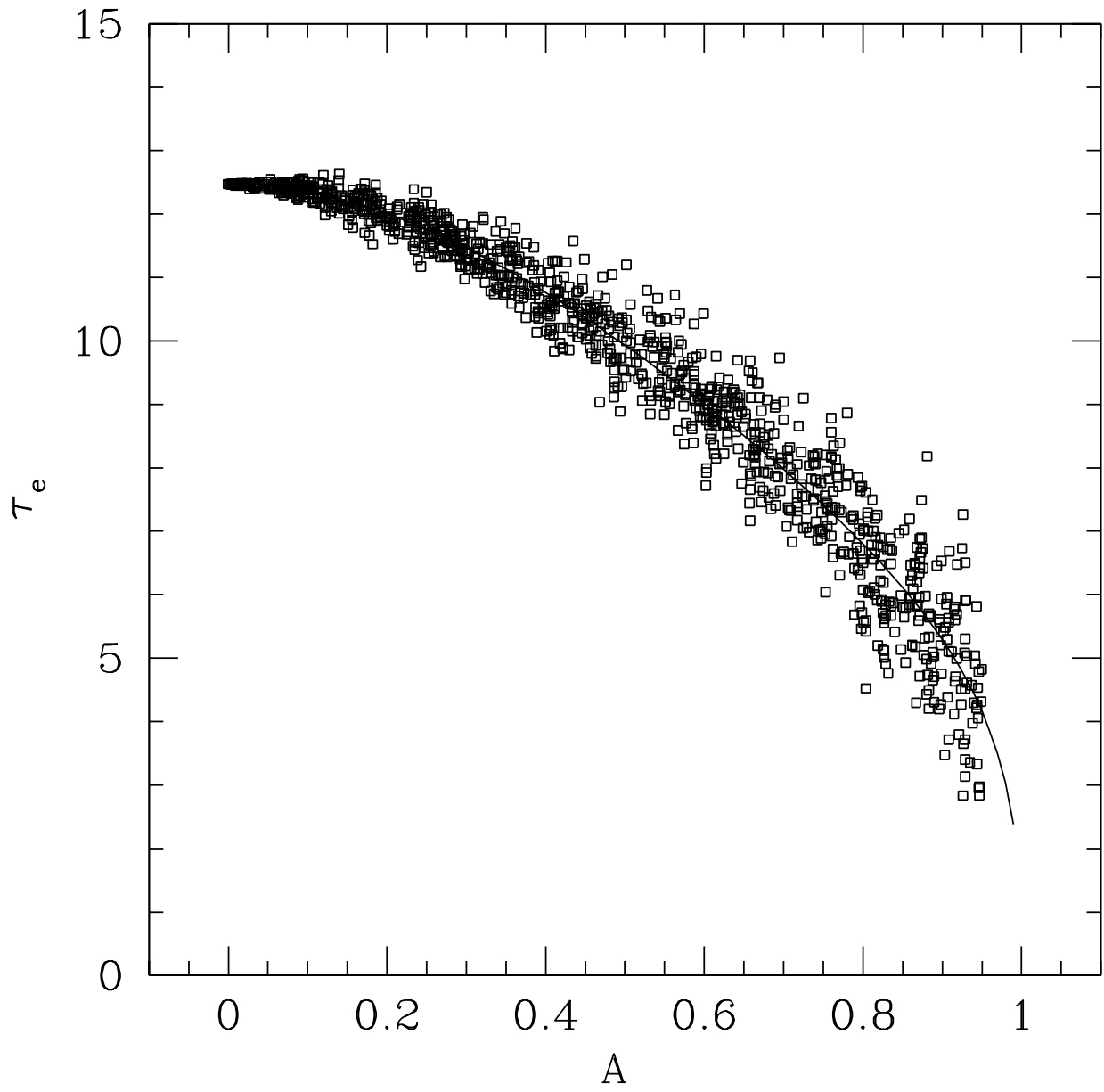
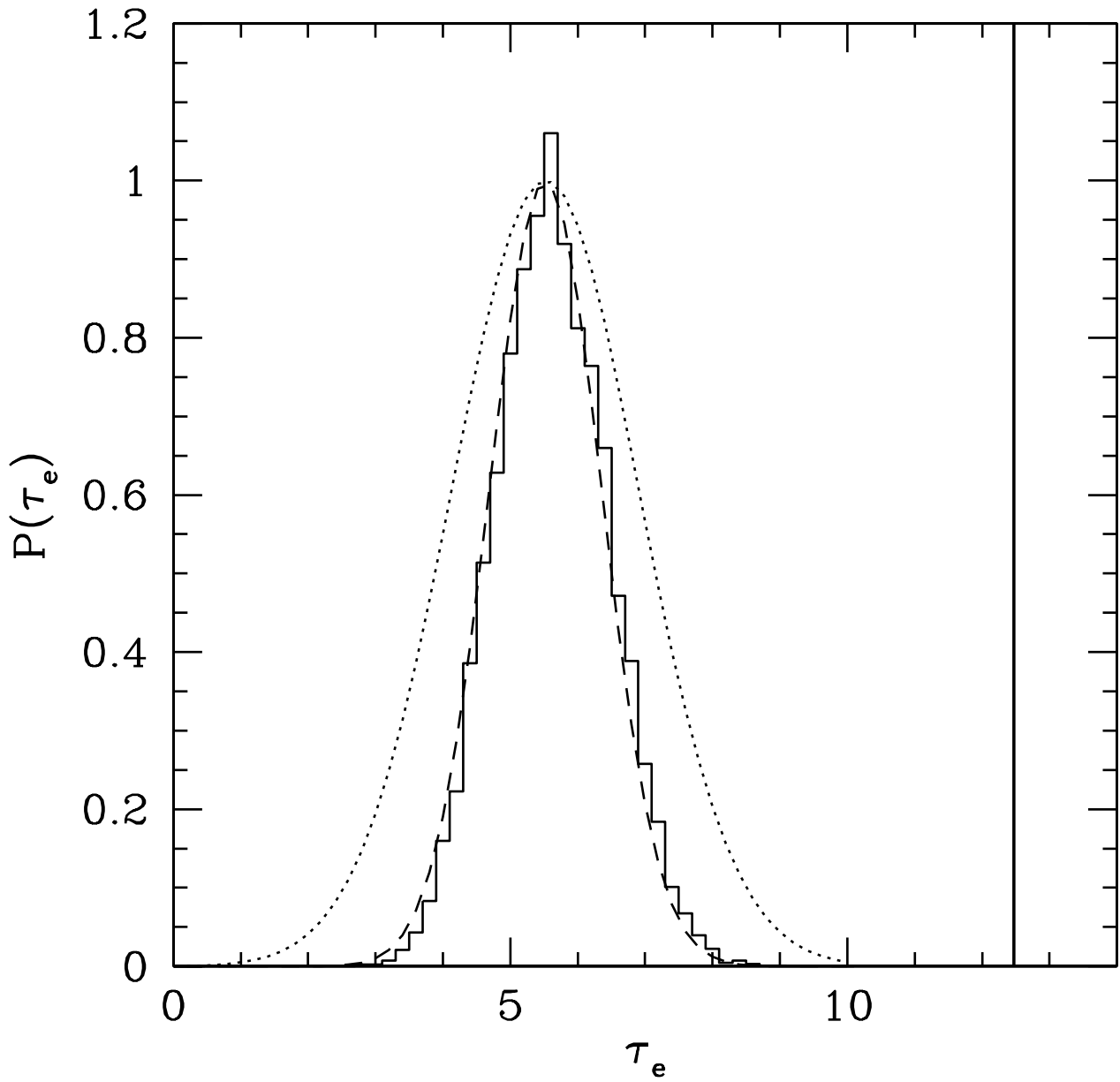
\end{document}